\documentclass[11pt,english]{article}
\usepackage{graphicx}
\usepackage{amstext}
\usepackage{caption}
\usepackage{etoolbox}
\usepackage{makeidx}
\include{epsf}
\usepackage{amsfonts}
\usepackage{authblk} 
\usepackage{sectsty}
\usepackage{amsmath,amssymb,epsfig}
\usepackage{amscd}
\usepackage{amsthm}
\usepackage{mathrsfs}
\usepackage{dsfont}
\usepackage[applemac]{inputenc}
\usepackage[english]{babel}
\usepackage{enumitem} 
\usepackage[]{latexsym}
\usepackage{caption}
\usepackage{subfigure}
\usepackage{hyperref}
\usepackage{blindtext}
\usepackage{verbatim}
\usepackage{cancel}

\hypersetup{
pdftitle={},%
pdfauthor={},%
pdfsubject={},%
pdfkeywords={},%
colorlinks=true,%
linkcolor=blue,%
citecolor=red,%
linktocpage=true,%
hyperfootnotes=true,%
pageanchor=true
}

\newcommand*{\affaddr}[1]{#1}
\newcommand*{\affmark}[1][*]{\textsuperscript{#1}}

\newtheorem*{proof*}{Proof}

\newcommand{\be}{\begin{equation}}

\newcommand{\ee}{\end{equation}}
\def\beqa{\begin{eqnarray}}
\def\eeqa{\end{eqnarray}}
\def\bean{\begin{eqnarray*}}
\def\eean{\end{eqnarray*}}

\renewenvironment{thebibliography}[1]
         {\section*{References}\frenchspacing\small
          \begin{list}{[\arabic{enumi}]}
         {\usecounter{enumi}\parsep=2pt\topsep 0pt
         \settowidth{\labelwidth}{[#1]}
         \leftmargin=\labelwidth\advance\leftmargin\labelsep
         \rightmargin=0pt\itemsep=1pt\sloppy}}{\end{list}}

 \numberwithin{equation}{section}

\input xy
\xyoption{all}

\usepackage[normalem]{ulem}

\newcommand{\ba}{\begin{eqnarray}}
\newcommand{\ea}{\end{eqnarray}}

\title{\textbf{\textsf{A note on the Hamiltonian as a polymerisation parameter}}\vspace{0.35cm}}

\author{
\textsf{Norbert Bodendorfer\affmark[1]\footnote{\texttt{norbert.bodendorfer@physik.uni-r.de}}, Fabio M. Mele\affmark[1]\footnote{\texttt{fabio.mele@physik.uni-r.de}}, and Johannes M\"unch\affmark[1]\footnote{\texttt{johannes.muench@physik.uni-r.de}}}\\
\affaddr{\affmark[1]\textsf{Institute for Theoretical Physics, University of Regensburg,}}\\
\affaddr{\textsf{93040 Regensburg, Germany}}\vspace{-0.5cm}
}

\begin{document}

\maketitle

\begin{abstract}
\textsf{In effective models of loop quantum gravity, the onset of quantum effects is controlled by a so-called polymerisation scale. It is sometimes necessary to make this scale phase space dependent in order to obtain sensible physics. A particularly interesting choice recently used to study quantum corrected black hole spacetimes takes the generator of time translations itself to set the scale. We review this idea, point out errors in recent treatments, and show how to fix them in principle.}
\end{abstract}

\section{Introduction}

Effective models of loop quantum gravity can be obtained from classical systems via certain substitutions that capture some of the quantum effects present in those systems. We focus here on so-called holonomy corrections, which are the analogues of approximating field strengths in terms of holonomies around plaquettes in lattice gauge theory.  As an example, one may substitute (polymerise) in cosmological models $b \mapsto \frac{\sin(\delta b)}{\delta}$ in the Hamiltonian, where $b$ is the Hubble rate. This introduces corrections to the equations of motion which are suppressed by powers of $\delta^2$ which is related to the Planck scale. For the simplest models, see. e.g. \cite{AshtekarRobustnessOfKey}, the choice $\delta \approx 1$ in natural units leads to phenomenologically viable scenarios with cosmological bounces at Planckian energy density. Different polymerisations based on different choices of variables may lead to a phase space dependent $\delta$ to obtain similar results, see. e.g. \cite{AshtekarQuantumNatureOfAnalytical}. In general, such phase space dependencies may greatly complicate the equations of motion and prohibit an analytic treatment. 

A middle ground between those two extremes (generic and constant $\delta$) is obtained for $\delta$ being a function of a constant of motion or the generator of time translations itself. In this case, $\delta$ remains constant along dynamical trajectories and the equations of motion may remain sufficiently simple. This approach was recently advocated in \cite{CorichiLoopquantizationof,OlmedoFromBlackHoles, AshtekarQuantumTransfigurarationof, AshtekarQuantumExtensionOf}. While the idea is quite interesting and seems to lead to physically sensible outcomes \cite{AshtekarQuantumTransfigurarationof, AshtekarQuantumExtensionOf}, one needs to be careful about the correct equations of motion ensuing from such choices. This note will comment on this topic and point out errors in recent treatments.

\section{The Hamiltonian as a polymerisation scale}

\subsection{Main idea and standard Hamiltonian formalism}

We consider the phase space $\Gamma = \mathbb R^{2n}$ with configuration variables $q^i$ and conjugate momenta $p_i$, $i=1,\ldots,n$. The Hamiltonian is of the form
\be
	H_{0} =  O(q^i, p_i, \delta) \text{,} \label{eq:HamDeltaFixed}
\ee
where a priori $\delta$ is a constant on $\Gamma$. $\delta$ plays the role of the polymerisation scale. We denote the Hamiltonian vector field of this Hamiltonian as $\vec v_{H_0}$.

In this paper, we consider the case when $\delta = f(O)$ is a function of the generator of time translations, i.e. the Hamiltonian. The Hamimiltonian is then recursively defined as 
\be
	H =  O(q^i, p_i, f(O)) \text{.} \label{eq:HamDeltaPhaseSpace}
\ee
The equations of motion for $q^i$ follow as
\ba
	\dot q^i &=& \frac{\partial H}{\partial p_i} =  \frac{\partial O}{\partial p_i} +  \frac{\partial O}{\partial \delta } \left( \frac{ \partial f}{\partial O} \left(  \frac{ \partial O}{\partial p_i} +  \frac{ \partial O}{\partial \delta } \left( \ldots \right) \right) \right) \nonumber \\
		   &=&  \frac{\partial O}{\partial p_i} \sum_{n=0}^\infty \left(  \frac{\partial O}{\partial \delta }  \frac{ \partial f}{\partial O} \right)^n \nonumber \\
		   &=&  \frac{1}{1-\frac{\partial O}{\partial \delta }  \frac{ \partial f}{\partial O}} \frac{\partial O}{\partial p_i} \label{eq:RescHamVec}
\ea
under the assumption $\left| \frac{\partial O}{\partial \delta} \frac{ \partial f}{\partial O}\right|  <1 $ and similarly for $p_i$. The dependence of $H$ on $q^i, p_i$ via the recursive $f(O)$ terms is hereby considered to be explicit. We emphasise that $\frac{\partial O}{\partial \delta} \frac{ \partial f}{\partial O} \neq 1$ in general despite the suggestive notation as $f(O)$ and $O(\delta)$ are in general unrelated functions. This in particular means that the Hamiltonian vector field gets rescaled by a phase space dependent factor as
\be
	\vec v_{H} = \frac{1}{1-\frac{\partial O}{\partial \delta }  \frac{ \partial f}{\partial O}}  \vec v_{H_0} \text{.} \label{eq:HamVecRes}
\ee
Violation of the condition $\left| \frac{\partial O}{\partial \delta}\frac{ \partial f}{\partial O} \right| <1$ signals that the solutions to the equations may behave in a singular fashion. A case by base study is  then necessary to determine whether a sensible and unique solution exists.

Let us discuss the consequences for the physics described by \eqref{eq:HamDeltaPhaseSpace} instead of \eqref{eq:HamDeltaFixed}. 
Constants of motion $D_k$ are not affected by this rescaling, i.e. they are selected both by \eqref{eq:HamDeltaFixed} and by \eqref{eq:HamDeltaPhaseSpace}. However, the local rescaling changes the time that passes between two physical events: comparing to the evolution generated by \eqref{eq:HamDeltaFixed} for a time $t$ when $\delta$ is set to the value $f(H)$ to an evolution generated by \eqref{eq:HamDeltaPhaseSpace} with time $\tilde t$, we must choose 
\be
	 t (\tilde t) = \int_{0}^{\tilde t} d \tilde t ' \frac{1}{1-\frac{\partial O}{\partial \delta }  \frac{ \partial f}{\partial O}} \label{eq:TimeChange}
\ee
to end up at the same phase space point. 

This observation allows for the following strategy to solve the equations of motion of the seemingly more complicated Hamiltonian \eqref{eq:HamDeltaPhaseSpace}. We first solve the equations of motion for \eqref{eq:HamDeltaFixed} and then insert \eqref{eq:TimeChange} to obtain the flow w.r.t. to the natural time of \eqref{eq:HamDeltaPhaseSpace}.

\subsection{Constrained formalism} \label{sec:Constrained}

An alternative derivation of the equations of motion can be attempted using constraints. 
We include this here to explain an error in the literature. 

Instead of specifying \eqref{eq:HamDeltaPhaseSpace} directly, we enlarge the phase space by $\delta$ and $p_\delta$ with $\{\delta, p_\delta\}=1$ following \cite{AshtekarQuantumExtensionOf} and define the total Hamiltonian
\be
	H_T =  O(q^i, p_i, \delta)  + \lambda \Phi, ~~~~~~~ \Phi :=   \delta - f(O(q^i, p_i, \delta))  \label{eq:HamTotal} \text{.}
\ee
$\Phi$ is a first class constraint and commutes with the Hamiltonian. We may now gauge fix \mbox{$\chi = p_\delta - h(q^i, p_i, \delta) = 0$}. Stability of this gauge fixing requires to fix $\lambda$ and leads to 
\be
	H_T =  O(q^i, p_i, \delta)  -  \frac{\frac{\partial O}{\partial \delta }+\{h,O\}}{1-\frac{\partial O}{\partial \delta }  \frac{ \partial f}{\partial O} - \{ h, f\} }\Phi   \label{eq:HamGaugeFixed} \text{.}
\ee
For $\chi$ to be a gauge fixing, it must be a second class constraint, i.e. $\{\chi,\Phi\} \neq 0$, which guarantees that the denominator in the second term of \eqref{eq:HamGaugeFixed} is non-zero. 
The gauge fixed Hamiltonian now depends explicitly on $h$, i.e. it is gauge dependent through the value of $\lambda$. One may also not argue that this gauge dependence is multiplied by $\Phi$, as we cannot use constraints before evaluating all Poisson brackets\footnote{Unless we use the Dirac bracket, in which case one again obtains the same results via $\{q^i, \delta\}_* \neq 0$ in general.}. A specific choice of $h$ satisfying $\{h, O\}\approx 0$ leads (modulo constraints) to 
\ba
	\dot q^i = \{ q^i , H_T\}  &= & \frac {\partial O(q^i, p_i, \delta)}{\partial p_i}  -  \frac{\frac{\partial O}{\partial \delta }}{1-\frac{\partial O}{\partial \delta }  \frac{ \partial f}{\partial O} } \left( -\frac{\partial f}{\partial p_i} \right)   \nonumber \\
					  & = &  \frac{\frac{\partial O}{\partial p_i} - \cancel{ \frac{\partial O}{\partial p_i} \frac{\partial f}{\partial O } \frac{\partial O}{\partial \delta} } + \cancel{ \frac{\partial O}{\partial \delta } \frac{\partial f}{\partial p_i } } }{1-\frac{\partial O}{\partial \delta }  \frac{ \partial f}{\partial O} } \nonumber \\
					  &=&  \frac{1}{1-\frac{\partial O}{\partial \delta }  \frac{ \partial f}{\partial O}} \frac{\partial O}{\partial p_i} 	
\ea
and thus to the expected equations of motion. Similarly, $q^i$ and $p_i$ are not Dirac observables w.r.t. $\Phi$ in general. One therefore should not expect them to be independent of the gauge choice $h$, which makes this line of derivation subtle\footnote{It is instructive to go through the simpler case $\delta = f(q^i, p_i)$, where $f$ is independent of $O$, where one also concludes that the gauge fixings leading to the correct result for the (gauge dependent) equations of motion is $h=h(O)$.}. 
On the other hand, constants of motion $D_k$ w.r.t. \eqref{eq:HamDeltaFixed} that do not depend on $p_\delta$ are Dirac observables w.r.t. $\Phi$. They are still constant w.r.t. the equations of motion generated by \eqref{eq:HamGaugeFixed}, as the extra terms are proportional to $\{D_k, \Phi \} \propto \{D_k, O\} = 0$. 

In other words, due to the gauge dependence, the derivation in this subsection needs an arbitrary input, the gauge fixing condition, which generally leads to different results. Any proof based on it therefore needs to supply an argument for why a specific gauge fixing is used, e.g. by computing the equations of motion independently, as done in the previous subsection. Otherwise, the logic is circular because the result is implicitly assumed via the gauge fixing. For a generic gauge fixing the connection to the initial Hamiltonian \eqref{eq:HamDeltaPhaseSpace} is lost.

In order to perform a technically correct phase space extension, one can first introduce a suitable gauge condition, e.g. $p_\delta =0$, i.e. $h = 0$, as a second class constraint conjugate to $\Phi$. This removes the necessity of the old phase space coordinates, $q^i$ and $p_i$, to commute with $\Phi$ at this stage of the phase space extension. In a second step, one may remove the gauge condition $p_\delta=0$ via gauge unfixing \cite{VytheeswaranGaugeUnfixingIn}. This adds corrections in the form of a power series in $p_\delta$ to $q^i$ and $p_i$. These extra terms are set to zero once the gauge $p_\delta=0$ is chosen. However, they now allow to choose any other gauge, e.g. the one from \cite{AshtekarQuantumExtensionOf}, as $q^i$ and $p_i$ are now gauge invariant.

\subsection{Compound and relativistic systems}

If one is only interested in correlations between phase space functions that are independent of the choice of time, as e.g. in relativistic systems, one does not need to worry about the phase space dependent rescaling of the Hamiltonian vector field observed in \eqref{eq:RescHamVec}. Again, owing to examples in the literature, it is worthwhile to point out systems of the type
\be
	H = N \left(O_1 - O_2 \right) \label{eq:CoupledSystem}
\ee
as e.g. in \cite{AshtekarQuantumTransfigurarationof, AshtekarQuantumExtensionOf}. Here, we have two independent instances of the above systems (with in general different functional forms of $O_{1,2}$) that are coupled via the same lapse function $N$. While the above reasoning for relativistic systems is correct for each subsystem independently, it is not for the joint systems as the relative rescalings of the Hamiltonian vector field are different in general. It follows that one needs to include \eqref{eq:TimeChange} in the analysis.

 \section{Comments on the literature} \label{sec:comments}
 
In order to avoid confusion, we point out some erroneous statements in the recent literature concerning the topic of this note. They are tied to the generally incorrect statement that for certain choices of $\delta$ which are not constant on phase space, the equations of motion are the same as for $\delta$ constant on phase space\footnote{A Dirac observable is in general not constant on phase space, but only along dynamical trajectories.}. For example, this would imply that the phase space dependent rescaling of the Hamiltonian vector field present in \eqref{eq:HamVecRes} would not appear. 

The first preprint version of \cite{AshtekarQuantumTransfigurarationof, AshtekarQuantumExtensionOf} incorrectly asserts that this is due to $\delta$ being a Dirac observable w.r.t. to the Hamiltonian constraint with constant $\delta$. In fact, in this case the vector fields are not even parallel in general. The published (2nd preprint) version of \cite{AshtekarQuantumTransfigurarationof, AshtekarQuantumExtensionOf} suggests that the above wrong claim is due to an unspecified special property of the system under consideration. The error in the proof of this statement in \cite{AshtekarQuantumExtensionOf} is to use a gauge fixing $p_\delta = h$ which is not of the above class yielding the correct (gauge dependent) equations of motion. 

Given this situation, it is now of great interest to check whether combining \eqref{eq:TimeChange} with \eqref{eq:CoupledSystem} leads to similar conclusions as in \cite{AshtekarQuantumTransfigurarationof, AshtekarQuantumExtensionOf} when applied to that system (with the same choices for $\delta$).
To this aim let us consider the effective Hamiltonian given in \cite{AshtekarQuantumExtensionOf} (cfr. Eqs. (2.18) and (A1), (A2))
\be
H_{eff} = N \left(O_1 - O_2\right) \quad, \quad O_1 := -\frac{1}{2 \gamma} \left[ \frac{\sin (\delta_b b)}{\delta_b} + \frac{\gamma^2 \delta_b}{\sin (\delta_b b)} \right] \frac{p_b}{L_o} \quad , \quad O_2 := \left[ \frac{\sin (\delta_c c)}{\gamma L_o \delta_c} \right] p_c \;, \label{eq:HeffAOS}
\ee
\noindent
together with the relations between the polymerisation scales $\delta_b$, $\delta_c$ and the subsystems of the Hamiltonian $O_1$, $O_2$ respectively given by (cfr. Eqs. (4.6) and (A14) in \cite{AshtekarQuantumExtensionOf})

\be
\delta_b = \left(\frac{\sqrt{\Delta}}{\sqrt{2 \pi} \gamma^2 O_1}\right)^{\frac{1}{3}} =: f_b(O_1) \quad , \quad L_o \delta_c = \frac{1}{2} \left(\frac{\gamma \Delta^2}{4 \pi^2 O_2}\right)^{\frac{1}{3}} =: L_o f_c(O_2) \;.
\ee
\noindent
Let then $t_1$ be the time corresponding to the choice $N = F_b := 1-\frac{\partial O_1}{\partial \delta_b }  \frac{ \partial f_b}{\partial O_1}$ for which the solutions for the $b$-sector $b(t_1)$, $p_b(t_1)$ have the same functional form as in \cite{AshtekarQuantumExtensionOf} (cfr. Eqs. (2.23), (2.25)), and let $t_2$ be the time corresponding to $N = F_c := 1-\frac{\partial O_2}{\partial \delta_c }  \frac{ \partial f_c}{\partial O_2}$ for which the solutions for the $c$-sector $c(t_2)$, $p_c(t_2)$ have the same functional form as in \cite{AshtekarQuantumExtensionOf} (cfr. Eqs. (2.21), (2.22)). The explicit computation of the derivatives occurring in the definitions of $F_{b,c}$ yields
\begin{align}
F_b(t_1) &= 1- \frac{p_b(t_1)}{6 \gamma L_o m} \left( 1 - \frac{\gamma^2 \delta_b^2}{\sin(\delta_b b(t_1))^2} \right) \left( b(t_1) \cos(\delta_b b(t_1)) - \frac{\sin(\delta_b b(t_1))}{\delta_b} \right) \;, 
\notag
\\
F_c(t_2) &= 1 + \frac{p_c(t_2)}{3 \gamma m} \left( \frac{c(t_2) \cos(\delta_c c(t_2))}{L_o} - \frac{\sin(\delta_c c(t_2))}{L_o \delta_c} \right)\;,\label{F}
\end{align}
where $O_1 = O_2 = m$ has been used after evaluation of the derivatives. 
By means of the inverse mapping of \eqref{eq:TimeChange} we can relate both times via
\begin{align}
\mathcal{I}_b(t_1) := \int^{t_1} F_b(t_1') dt'_1 \quad&, \quad \mathcal{I}_c(t_2) := \int^{t_2} F_c(t_2') dt'_2 \; ,
\notag
\\
t_1(t_2) = \mathcal{I}_b^{-1}(\mathcal{I}_c(t_2)) \quad &, \quad t_2(t_1) =\mathcal{I}_c^{-1}(\mathcal{I}_b(t_1)) \;,
\end{align}
\noindent
which requires convergence of the integrals and invertibility of $\mathcal{I}_b(t_1)$ and $\mathcal{I}_c(t_2)$.
\begin{figure}[t!]
	\centering
	\subfigure[]
	{\includegraphics[width=7.75cm,height=5.5cm]{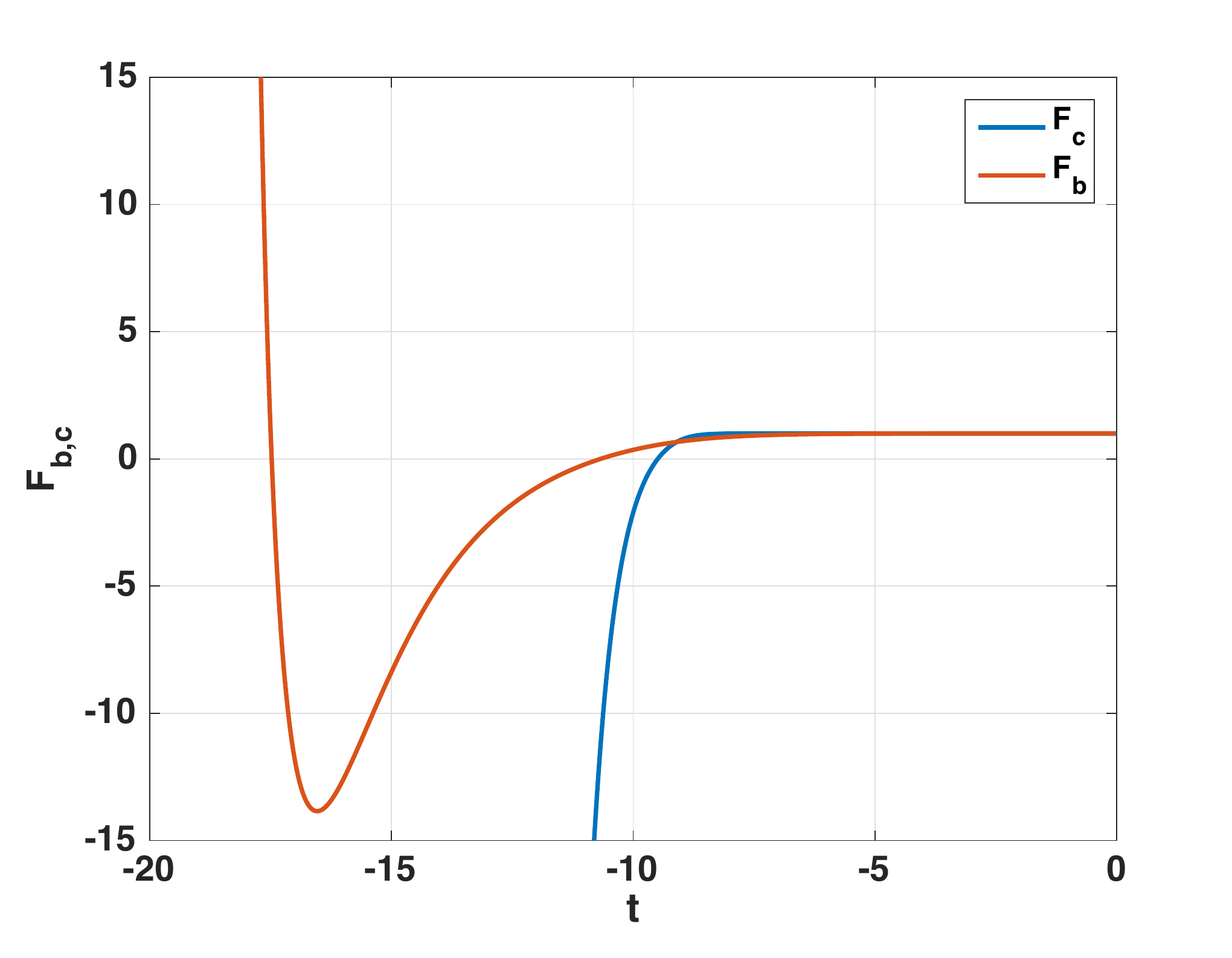}}
	\hspace{2mm}
	\subfigure[]
	{\includegraphics[width=7.75cm,height=5.5cm]{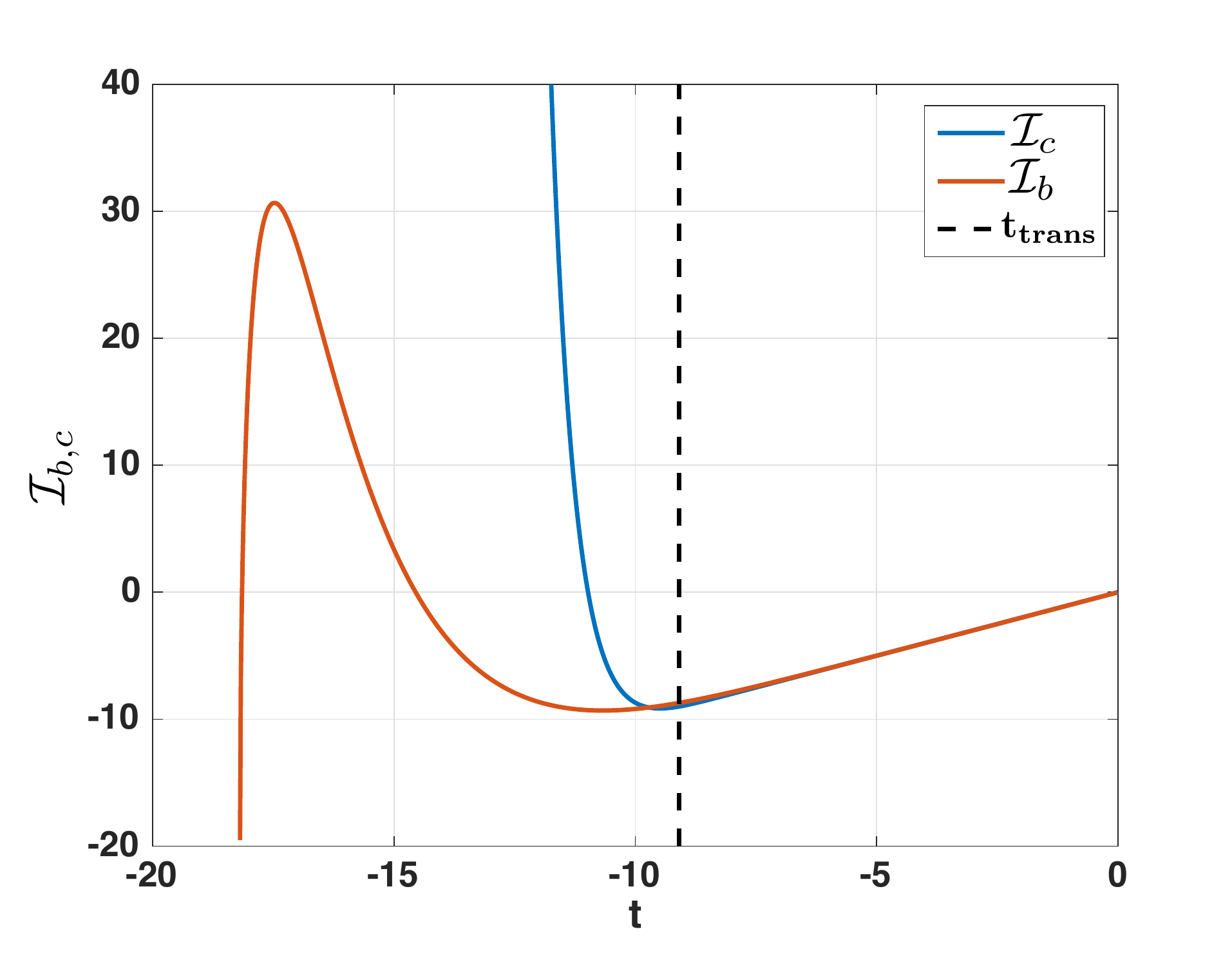}}
	\caption{a) shows $F_b$ and $F_c$ along the dynamical trajectories $b(t_1)$, $p_b(t_1)$ and $c(t_2)$, $p_c(t_2)$, respectively, while b) shows the numerical integration of $\mathcal{I}_b(t_1)$ and $\mathcal{I}_c(t_2)$ for the parameters $L_o = 1$, $\Delta = 1$, $m = 10000$, $\gamma = 0.2375$. The dashed black line denotes the $t_2$-value of the transition surface.}
	\label{figure}
\end{figure}
As shown in Fig. \ref{figure} a), one generically finds that $\left| \frac{\partial O}{\partial \delta}\frac{ \partial f}{\partial O} \right| = 1$ is crossed along dynamical trajectories (at different times in the two subsystems) in such a way that integrating the equations of motion leads to singular results. This also signals that the effective time flow as in \eqref{eq:TimeChange} changes direction along dynamical trajectories. Moreover, as showed in Fig. \ref{figure} b), $t_1(t_2) \simeq t_2$ close to the horizon ($t = 0$) up to the transition surface ($t = t_{trans}$) where the bounce occurs. This signals that the solutions derived in \cite{AshtekarQuantumTransfigurarationof, AshtekarQuantumExtensionOf} can be considered as a good approximation for the effective dynamics w.r.t. to the correct equations of motion derived from \eqref{eq:HeffAOS} only in the black hole interior region up to the transition surface. This is reasonable as corrections to the classical equations are small in this region. Soon after the transition surface however, when quantum effects have become large at least for some time, the system ceases to be have a well-defined Hamiltonian vector field as indicated in Fig. \ref{figure} a), i.e. the rescaling $1/F_{b,c}$ in \eqref{eq:HamVecRes} diverges. Deviations from the classical limit ($F_b \approx F_c \approx 1$) at late times ($t \ll 0$) are due to the derivatives $\partial O_{1,2}/\partial \delta_{b,c}$, which lead to ``naked connections'' $b(t_1), c(t_2)$ that transition from $\delta_c c(t_1), \delta_b b(t_2) \approx 0$ on the black hole side to $\delta_c c(t_1), \delta_b b(t_2) \approx\pi$ on the white hole side.

Similarly, the earlier paper \cite{OlmedoFromBlackHoles} also uses equations of motion where the local rescaling in \eqref{eq:HamVecRes} is absent, although the Hamiltonian has the form \eqref{eq:CoupledSystem} where $\delta$ is related to the mass of the black hole, which is the on-shell value of $O_{1,2}$. 

Despite this technical problem, the calculations in \cite{CorichiLoopquantizationof,OlmedoFromBlackHoles, AshtekarQuantumTransfigurarationof, AshtekarQuantumExtensionOf} are still very interesting and provide important steps towards a better understanding of black hole spacetimes in loop quantum gravity. One can put these calculations on a sound conceptual footing by simply postulating the equations of motion without the rescaling in \eqref{eq:HamVecRes} and drop the claim that they follow from an effective Hamiltonian. It may still be the case that such theories follow in a suitable limit from the full quantum theory, as they capture holonomy-type corrections. Alternatively, one may take the extended phase space from section \ref{sec:Constrained} as a definition of the system and consider the gauge dependence of the equations of motion to be a model building feature instead of a bug.

We would like to stress that we do not claim that the equations of motion \eqref{eq:RescHamVec} obtained from the gauge choice $\{h,O\}\approx 0$ in the context of \cite{AshtekarQuantumTransfigurarationof, AshtekarQuantumExtensionOf} as discussed in section \ref{sec:comments} lead to physically more sensible results than the equations used in \cite{AshtekarQuantumTransfigurarationof, AshtekarQuantumExtensionOf}. But we also stress that they are the correct equations of motion derived from Hamiltonians of the type \eqref{eq:HamDeltaPhaseSpace} via Hamilton's equations.

\section*{Acknowledgments}

The authors were supported by an International Junior Research Group grant of the Elite Network of Bavaria. The authors would like to thank Abhay Ashtekar, Javier Olmedo, and Parampreet Singh for extensive email discussions about their work.


\begin{thebibliography}{1}
	
	\bibitem{AshtekarRobustnessOfKey}
	A.~Ashtekar, A.~Corichi, and P.~Singh, ``{Robustness of key features of loop
		quantum cosmology},'' {\em Physical Review D} {\bf 77} (2008) 024046, {\tt
		arXiv:0710.3565 [gr-qc]}.
	
	\bibitem{AshtekarQuantumNatureOfAnalytical}
	A.~Ashtekar, T.~Pawlowski, and P.~Singh, ``{Quantum nature of the big bang: An
		analytical and numerical investigation},'' {\em Physical Review D} {\bf 73}
	(2006) 124038, {\tt arXiv:gr-qc/0604013}.
	
	\bibitem{CorichiLoopquantizationof}
	A.~Corichi and P.~Singh, ``{Loop quantization of the Schwarzschild interior
		revisited},'' {\em Classical and Quantum Gravity} {\bf 33} (2016) 055006,
	{\tt arXiv:1506.08015 [gr-qc]}.
	
	\bibitem{OlmedoFromBlackHoles}
	J.~Olmedo, S.~Saini, and P.~Singh, ``{From black holes to white holes: a
		quantum gravitational, symmetric bounce},'' {\tt arXiv:1707.07333 [gr-qc]}.
	
	\bibitem{AshtekarQuantumTransfigurarationof}
	A.~Ashtekar, J.~Olmedo, and P.~Singh, ``{Quantum Transfiguration of Kruskal
		Black Holes},'' {\em Physical Review Letters} {\bf 121} (2018) 241301, {\tt
		arXiv:1806.00648 [gr-qc]}.
	
	\bibitem{AshtekarQuantumExtensionOf}
	A.~Ashtekar, J.~Olmedo, and P.~Singh, ``{Quantum extension of the Kruskal
		spacetime},'' {\em Physical Review D} {\bf 98} (2018) 126003, {\tt
		arXiv:1806.02406 [gr-qc]}.
	
	\bibitem{VytheeswaranGaugeUnfixingIn}
	A.~S. Vytheeswaran, ``{Gauge unfixing in second-class constrained systems},''
	{\em Annals of Physics} {\bf 236} (1994), no.~2 297--324.
	
\end{thebibliography}
\end{document}